\documentclass[12pt,preprint]{aastex}

\newcommand{\um}{$\mu$m~}
\newcommand{\ums}{$\mu$m}

\shorttitle{Infrared Stellar Excesses in $Spitzer$ Surveys}
\shortauthors{Hovhannisyan et al.}

\begin{document}

\title{$Spitzer$ 24\,\um Excesses for Bright Galactic Stars in Bo\"{o}tes and First Look Survey Fields}

\author{L.R. Hovhannisyan\altaffilmark{1}, A.M. Mickaelian\altaffilmark{1}, D.W. Weedman\altaffilmark{2},  E. Le Floc'h\altaffilmark{3}, J.R. Houck\altaffilmark{2}, B.T. Soifer\altaffilmark{4,5},  K. Brand\altaffilmark{6}, A. Dey\altaffilmark{7}, B.T. Jannuzi\altaffilmark{7} }

\altaffiltext{1}{Byurakan Astrophysical Observatory and Isaac Newton Institute of Chile Armenian Branch, 378433, Byurakan, Aragatzotn Province, Armenia; lilithov11@yahoo.com}
\altaffiltext{2}{Astronomy Department, Cornell University, Ithaca, NY 14853; dweedman@isc.astro.cornell.edu}
\altaffiltext{3}{Steward Observatory, University of Arizona, Tucson, AZ, 85721, and $Spitzer$ Fellow, Institute for Astronomy, University of Hawaii, Honolulu, HI, 96822}
\altaffiltext{4}{Spitzer Science Center, California Institute of Technology, 220-6, Pasadena, CA 91125}
\altaffiltext{5}{Division of Physics, Mathematics and Astronomy, 320-47, California Institute of Technology, Pasadena, CA 91125}
\altaffiltext{6}{Space Telescope Science Institute, 3700 San Martin Dr., Baltimore, MD, 21218}
\altaffiltext{7}{National Optical Astronomy Observatory, 950 N. Cherry Ave.,
Tucson, AZ, 85719}

\begin{abstract}
  
Optically bright Galactic stars ($V$ $\la$ 13 mag) having f$_{\nu}$(24 \ums) $>$ 1\,mJy are identified in $Spitzer$ mid-infrared surveys within 8.2 deg$^{2}$ for the Bo\"{o}tes field of the NOAO Deep Wide-Field Survey and within 5.5 deg$^{2}$ for the First Look Survey (FLS).  128 stars are identified in Bo\"{o}tes and 140 in the FLS, and their photometry is given.  ($K$-[24]) colors are determined using $K$ magnitudes from the 2MASS survey for all stars in order to search for excess 24\,\um luminosity compared to that arising from the stellar photosphere.  Of the combined sample of 268 stars, 141 are of spectral types F, G, or K, and 17 of these 141 stars have 24 \um excesses with ($K$-[24]) $>$ 0.2 mag.  Using limits on absolute magnitude derived from proper motions, at least 8 of the FGK stars with excesses are main sequence stars, and estimates derived from the distribution of apparent magnitudes indicate that all 17 are main sequence stars.  These estimates lead to the conclusion that between 9\% and 17\% of the main sequence FGK field stars in these samples have 24 \um infrared excesses.  This result is statistically similar to the fraction of stars with debris disks found among previous $Spitzer$ targeted observations of much brighter, main sequence field stars. 
\end{abstract}



\section {Introduction}

The sensitivity and efficiency of the Spitzer Space Telescope ($Spitzer$) have enabled numerous wide-area surveys with the Multiband Imaging Photometer for $Spitzer$ (MIPS, Rieke et al. 2004) to seek mid infrared sources with f$_{\nu}$(24 \ums) $>$ 0.3 mJy.  Such objects are often unusually bright in the mid infrared compared to other spectral regions.  At high Galactic latitudes, these surveys include the NOAO Deep Wide Field Survey (NDWFS) in Bo\"{o}tes \citep{jan99,hou05,bra06}, the Spitzer Wide-area Infrared Extragalactic Survey (SWIRE, Lonsdale et al. 2003, 2004), and the Spitzer First Look Survey (FLS, Fadda et al. 2006). To date, these surveys have been used only to locate unusual extragalactic sources, primarily dusty star forming galaxies and active galactic nuclei of very high luminosity \citep[e.g.][]{wee06,yan07,hou07,wee08}.  

These surveys also include numerous bright, Galactic stars.  It is known from targeted observations with $Spitzer$ that some stars have mid-infrared excesses at 24 \ums, usually attributed to circumstellar dust arising in debris disks \citep {gor06, bry06, rie05, su06, mey08}.  By using similar criteria for defining infrared excess as in these previous targeted studies, it is possible to use the wide area $Spitzer$ surveys to produce unbiased lists of stars with infrared excesses.  The high latitude surveys are especially useful, because they minimize confusion with other Galactic stars and are in regions chosen to have low zodiacal background.  

In the present paper, we identify and discuss the characteristics of Galactic stars in the Bo\"{o}tes and FLS survey areas (the first two surveys for which we had access to complete and reliable catalogs), giving in total a flux limited sample of Galactic stars within an area of $\sim$ 14 deg$^{2}$. We list all stellar sources (128 in Bootes and 140 in the FLS) with $r$ $<$ 17 mag. and f$_{\nu}$(24 \ums) $>$ 1 mJy.  Among these stars, we identify those which have excess f$_{\nu}$(24 \ums) compared to that expected from a stellar photosphere. 

For comparison to the f$_{\nu}$(24 \ums), we utilize available $J$, $H$, $K$ photometry from the Two Micron All Sky Survey (2MASS, Skrutskie et al. 2006).  (In this paper, we use $K$ to mean the 2MASS K$_{s}$ magnitude.) As a criterion for excess, we use the ($K$-[24]) color, as in \citet{gor04} and \citet{gor06}, and use previous targeted studies to compare the ($K$-[24]) colors with those for our samples.  Images and $r$ band photometry from the Sloan Digital Sky Survey (SDSS, Gunn et al. 1998, Adelman-McCarthy et al. 2008) are used to define the optical magnitudes; proper motions and $V$ photometry from the Tycho mission \citep{hog00}, and spectral classifications from the Digitized First Byurakan Survey (DFBS, Mickaelian et al. 2007\footnote {http://byurakan.phys.uniroma1.it/}) are also used to aid in estimating the luminosity classes of the stars.  Images from the Digitized Sky Survey (DSS) are also used in determining the morphological classification of sources (stars or galaxies)\footnote {http://stdatu.stsci.edu/cgi-bin/dss}.

\section{Observations}
\subsection{Bo\"{o}tes and FLS Surveys}

The Bo\"{o}tes MIPS survey was obtained for the 8.2 deg$^{2}$ within the Bo\"{o}tes field of the NOAO Deep Wide-Field Survey.  This survey was done as part of $Spitzer$ Guaranteed Time Observations by J. Houck and M. Rieke, and a catalog of 24 \um sources (not yet published) was produced by E. Le Floc'h.  The MIPS survey had an effective integration time at 24\,\um of $\sim$ 90\,s per sky pixel, reaching a 5 $\sigma$ detection limit of $\sim$ 0.3\,mJy for unresolved sources.  The FLS has similar integration times and detection limits for the full survey within 5.5 deg$^{2}$ \citep{fad06}.  Uncertainties are quoted in the FLS catalog for each source; for sources with f$_{\nu}$(24 \ums) $\ga$ 1 mJy, 5 $\sigma$ uncertainties range from 0.15\,mJy to 0.3\,mJy.  

For the present study, we restrict our analysis to sources in Bo\"{o}tes and the FLS having f$_{\nu}$(24 \ums) $>$ 1 mJy to assure that the infrared detections are of high precision, with S/N $>$ 15. The 5 $\sigma$ positional uncertainty for unresolved sources with these flux densities is 2.0\arcsec, based on comparisons of 24 \um sources with unambiguous optical identifications. 

The Bo\"{o}tes and FLS surveys use much deeper MIPS observations than in the targeted studies of bright stars distributed over the sky such as in \citet{bry06} and \citet{mey08}.  The optical/near-infrared magnitudes and the f$_{\nu}$(24 \ums) of these targeted field stars are much brighter than for the Bo\"{o}tes and FLS stars.  For example, the average $J$ magnitude of the stars in \citet{bry06} is 4.42, whereas for the stars in Bo\"{o}tes and the FLS is 9.13, a factor of 77 fainter.  

Despite the fact that we are observing much fainter stars than the targeted studies, we are sensitive to a similar fraction of infrared excess above the expected photospheric emission.  This is because our 24\,\um observations reach fainter relative to the expected photospheric emission than do the 24\,\um observations for the targeted field stars.  For example, the average photospheric emission at 24\,\um for the Bryden et al. sample is 260 mJy, and the criterion for excess in that sample was an excess of 20\% compared to the flux density expected from the photosphere.  Stars in our sample, being a factor of 77 fainter, should have average photospheric emission at 24\,\um of 3.4 mJy.  The 3$\sigma$ flux density uncertainties for our detections are $\la$ 0.2 mJy, which means that we could detect infrared excesses $\ga$ 10\% of the expected photospheric emission.


\subsection {Identification of Stars in Bo\"{o}tes and FLS Fields}

In order to identify bright Galactic stars, our objective was to identify all optically-unresolved sources with SDSS $r$ $<$ 17 and f$_{\nu}$(24 \ums) $>$ 1 mJy in the Bo\"{o}tes and FLS surveys.  The optical magnitude limit is chosen to assure that sources are sufficiently bright that they are Galactic stars rather than extragalactic quasars. (One known quasar was rediscovered, and has been excluded from the list.) The requirement that sources be optically unresolved is to assure that we are not selecting galaxies. 

To identify bright stars, we first generated lists of $Spitzer$ 24\,\um sources which are identified with optical sources in the survey catalogs, using as the criterion for identification that coordinates agree to within 2.0\arcsec~ in the merged infrared/optical catalogs.  Our initial criteria for identifying sources optically used $R$ mag. because these are the magnitudes given in the merged catalogs.  The NDWFS and FLS surveys are designed primarily for the identification of very faint optical sources, to $R$ $\sim$ 26 mag., and the positional identification criterion works well for faint, unresolved sources.  Brighter, resolved galaxies and stars brighter than $R$ $\sim$ 16 may not be reliably identified in this way because of such problems as saturation of the source image or complex image structure.  It is sometimes difficult, for example, to determine if saturated sources in the optical survey images are stars or galaxies.  

The best way to assure that all optically bright sources have been identified is to examine individually the location of each 24\,\um source which does not have an optical counterpart in the merged catalog in order to determine if the absence of a counterpart is because the optical source is too faint, or because it is too bright and too large on the images for an accurate position.  To assure a complete list of bright stars, it is also necessary to examine individually all sources with listed counterpart magnitudes $R$ $<$ 17 to verify if these sources are stars or galaxies. The magnitude cutoff of $R$ $<$ 17 is chosen to assure that all bright Galactic stars are included but to exclude extragalactic quasars and unresolved galaxies which appear in increasing numbers for $R$ $>$ 17. 

Even though $R$ magnitudes had to be used to define the initial search for sources, because these are the magnitudes tabulated in the catalogs, $R$ photometry in the Bo\"{o}tes and FLS surveys is unreliable for many of the brighter stars ($R$ $\la$ 15) because the images are saturated. In order to have consistent and reliable magnitudes over the full range of magnitudes selected, our final tabulation of sources uses SDSS $r$ mag.  Inclusion of a source in our final list as a Galactic star (which we also call "stellar" objects below) requires SDSS $r$ $<$ 17, requires that the source be unresolved in optical images, and requires that the source not be included in any existing list of quasars.  

In the Bo\"{o}tes survey area, there are 2853 sources with f$_{\nu}$(24 \ums) $>$ 1 mJy.  Of these, 308 sources have no optical counterpart in the merged catalog within 2.0\arcsec, and 281 sources have a counterpart identified which has $R$ $<$ 17.   We used several steps to identify and classify these 589 sources. In the first step, DSS1 and DSS2 red and blue images were examined to classify the objects morphologically.  Sources were also examined on the SDSS images\footnote {http://cas.sdss.org/astrodr6/en/tools/crossid/upload.asp} for confirmation of the classification and to obtain SDSS $r$ magnitudes.  Unresolved sources are readily separated from galaxies in these optical images because of the presence of diffraction spikes for unresolved sources.

Of the 589 sources within Bo\"{o}tes selected for initial examination of their images, 439 were found to be galaxies bright enough for classification on either DSS or SDSS images, 16 were found to be sources too faint for classification on the DSS or SDSS ($r$ $>>$ 17), 6 were optically unresolved objects with $r$ $>$ 17, and 128 were optically unresolved objects with $r$ $<$ 17 which are not known quasars.  These 128 sources are listed in Table 1 as Galactic stars.  Despite using the $r$ $<$ 17 criterion for inclusion in the sample, all stars in Table 1 are actually $r$ $<$ 16.  This is simply an empirical result; there are no stellar sources in the survey with 16.0 $<$ $r$ $<$ 17.0, which confirms that our criterion of $r$ $<$ 17 is a relaxed criterion that assures inclusion of all bright Galactic stars.  

The FLS survey catalog lists f$_{\nu}$(24 \ums) and $R$ magnitudes for sources.  We used a similar strategy as in Bo\"{o}tes to examine all sources which could be bright stars.   All sources in the catalog having f$_{\nu}$(24 \ums) $>$ 1 mJy and with listed $R$ $<$ 17 or having no entry for $R$ were examined individually on DSS and SDSS images to determine if they are bright Galactic stars, or galaxies.  In the FLS area, there are 213 sources having f$_{\nu}$(24 \ums) $>$ 1 mJy with an optical counterpart $R$ $<$ 17 and 87 sources with no optical counterpart.  Of these 300 sources, 140 satisfy the criteria described above for being classified as Galactic stars.  As for Bo\"{o}tes, our final selection of sources uses SDSS $r$ $<$ 17 mag. to provide consistent photometry which extends to brighter stars.  The 140 stellar sources in the FLS are listed in Table 2.  Despite using the $r$ $<$ 17 criterion for inclusion in the sample, all stars are actually brighter; $r$ $<$ 15.5 for the FLS stars.

\subsection{Determination of 24\,\um Excesses}

The objective of our study is to identify excess infrared continuum emission which exceeds that expected from the stellar photosphere.  Determining the spectral energy distribution (SED) and temperature distribution of the material producing the infrared excess requires detailed modeling of the stellar atmospheres and subsequent measures of the deviations from such a model at various wavelengths \citep[e.g. ][]{bry06, mey06, mey08}. 

If the objective is only to identify stars with infrared excesses without determining in detail the spectrum of the excess, this identification can be done in a straightforward, empirical way with limited photometry.  As shown in \citet{gor04}, the identification of excesses can be done with confidence simply by comparing near-infrared photometry ($J$, $H$, $K$) with the f$_{\nu}$(24 \ums).  

This method for determining 24\,\um excesses in stars compares colors $J$-$K$ with colors $K$-[24]; magnitude at 24\,\um is defined as [24] with zero magnitude of the MIPS f$_{\nu}$(24 \ums) corresponding to 7300 mJy \citep{gor04}.  As pointed out by Gorlova et al., this method works because pure photospheric emission leads to $K$-[24] = 0, virtually independent of spectral type through type K.  This constant color arises because both the $K$ and [24] bands are on the Rayleigh-Jeans tail of the spectral energy distribution so that the $K$-[24] color for stars hotter than type M has little dependence on stellar temperature.  

\citet{gor04} define stars with 24\,\um excesses as those having a $>$ 3$\sigma$ excess at 24\,\um compared to the f$_{\nu}$(24 \ums) expected from the photosphere, with the value of $\sigma$ determined from the uncertainties of the 2MASS and MIPS photometry.  With these uncertainties, stars with ($K$-[24]) $>$ 0.2 mag can be identified as stars with probable excesses. The observations and stellar models in \citet{gau07} demonstrate, however, that M stars are sufficiently cool that they can show colors ($K$-[24]) $>$ 0.2 from photospheric continuum alone.  Although we identify such M stars in the following analysis, we do not include them as sources with real 24 \um excesses.  

The validity of this simple criterion can be tested using the stars identified in \citet{bry06} and \citet{mey08} as having excesses at 24 \um arising from debris disks, as determined from comparison of f$_{\nu}$(24 \ums) to the photospheric f$_{\nu}$(24 \ums) expected from flux densities at other wavelengths.  \citet{bry06} use multiwavelength fits of the SEDs from stellar models for 69 stars to determine excesses at 24 \um and 70 \um. \citet{mey06,mey08} and \citet{car08,car09} use stellar models to compare f$_{\nu}$(24 \ums) with the expected values from photospheric emission for 328 stars and list 30 stars which show excesses at 24 \um but no excesses at 8 \ums.  For all of the stars listed in \citet{bry06} and \citet{mey08}, we have taken the archival 2MASS data and compared to the 24 \um data in these papers.  

The distribution of $K$-[24] colors for the stars with excesses given in Meyer et al is shown in the lowest panel of Figure 1.  It can be seen from this plot that ($K$-[24]) $>$ 0.2 mag is a conservative definition of excess, because some of the stars with excesses have $K$-[24] as small as 0.15.  

For the results in \citet{bry06}, values of the 24 \um flux densities are given for all stars observed, so we show all of the stars in the ($J$-$K$), ($K$-[24]) color-color plot.  Based on photospheric modeling, Bryden et al identify only one star (HD 33262) as having an excess at 24 \ums.  For this star, ($K$-[24]) = 0.41, so it satisifies our color criterion for an excess.  In Figure 1, there are 3 additional stars with ($K$-[24]) $>$ 0.2 (HD 115383, HD 117176, HD 35296).  Of these 3, HD 117176 is identified by Bryden et al as having an excess at 70 \um, so we suggest that these 3 stars also have evidence for debris disks based on the ($K$-[24]) color.  In sum, therefore, there are 4 of 69 stars (6\%) with ($K$-[24]) $>$ 0.2 mag.

127 of our 128 stars in Bo\"{o}tes and 139 of 140 in the FLS have $J$, $H$, and K magnitudes from 2MASS, so the ($J$-$K$), ($K$-[24]) color-color comparison is easily done.  Results for the Bo\"{o}tes stars and FLS stars are also shown in Figure 1. Because 3$\sigma$ uncertainties in the Bo\"{o}tes and FLS data are $\la$ 0.2 mJy, we identify those stars with ($K$-[24]) $>$ 0.2 mag. as stars with possible excesses.  Using this definition of excess, there are 13 such stars in  Bo\"{o}tes and also 13 in the FLS.  While in some cases, the observed excesses may arise because of sources being at extremes of the error, this definition of excess is justified by noting in Figure 1 that there is more extended dispersion of the points with ($K$-[24]) $>$ 0 than for ($K$-[24]) $<$ 0.  If the observed excesses arise simply from a distribution of errors, the positive and negative distributions should be similar.   

The stars with infrared excesses defined by ($K$-[24]) $>$ 0.2 mag. are identified in Figure 1 according to their listing in Tables 1 and 2.  The most extreme is source 107 in Bo\"{o}tes, but this is a known variable star \citep{ak00}, so the observed excess may arise because the $K$ was different when [24] was measured than the $K$ at the 2MASS observing epoch.  In principle, variability could account for the apparent excess found in any star, but this is the only star with identified excess which is a known variable.  If variability does explain apparent 24 \um excesses, we would expect a similar number of variable stars which show apparent 24 \um "deficiencies" (i.e. $K$-[24] $<$ -0.2 mag.).  Of the total sample of 266 Bo\"{o}tes and FLS stars in Figure 1, only two show 24 \um deficiencies; this implies statistically that most of the observed excesses are real and not caused by source variability. 

Not counting the variable source 107, there remain 12 stars in Bo\"{o}tes and 13 in the FLS with 0.2 $<$ ($K$-[24]) $<$ 0.7.  These are the stars which we identify as candidates for having real infrared excesses at 24 \um and discuss further below in section 3. 

Except for the measurement uncertainty, there is only one other source of uncertainty regarding the reality of the measured excesses.  This is the question of whether the observed 24\,\um excess might arise from a faint, background galaxy which produces some of the f$_{\nu}$(24 \ums) but is not visible in optical or 2MASS images because of the bright star.  For the 26 stars with excesses, a comparison of the 24\,\um coordinates with the 2MASS coordinates yields a median difference of 0.7\,\arcsec~ between coordinates.  This means that any extragalactic source which confuses the f$_{\nu}$(24 \ums) has to be located precisely at the stellar position and not simply somewhere within the $\sim$ 5\,\arcsec~ beam size of the MIPS survey camera.  

The probability of such a coincidence by chance is extremely small.  We estimate this using the principle that a random source producing the apparent excess should have a minimum f$_{\nu}$(24 \ums) which is 20\% of the photospheric f$_{\nu}$(24 \ums) (in order to produce an apparent $K$-[24] excess of 0.2 mag.).  The faintest sources, f$_{\nu}$(24 \ums) $\sim$ 1 mJy, are most susceptible to such confusion, so we determine the probability that a faint source of f$_{\nu}$(24 \ums) $>$ 0.2 mJy is randomly located within $\sim$ 1\arcsec~ of a given position.  To f$_{\nu}$(24 \ums) = 0.2 mJy, there are $\sim$ 6x10$^{3}$ sources deg$^{-2}$ \citep{pap04}, so the probability of having a random source within an area of 1\arcsec~ radius is 1.4x10$^{-3}$.  For a 5 mJy star, requiring coincidental excess of 1 mJy, the probability of a 1 mJy random source is 8x10$^{-5}$ using our observed source density in Bo\"{o}tes of 350 sources deg$^{-2}$ to 1 mJy.

\section{Discussion} 

\subsection {Spectroscopic Classifications}

Previous targeted observations using $Spitzer$ have enabled extensive studies of circumstellar material associated with a wide variety of stars.  Objectives range from probing the formation and evolution of planetary systems (FEPS, e.g. Meyer et al. 2006) to understanding dust formation in evolved stars \citep[e.g. ][]{mat07}.  A primary goal of the FEPS studies is to determine the frequency and nature of debris disks surrounding stars similar in luminosity and age to the sun in order to place our solar system in context of other similar stars. Knowledge regarding the luminosity of such disks is also crucial in planning eventual efforts to image terrestrial planets because bright zodiacal emission associated with a planetary system is a fundamental limitation to such imaging \citep{bei06}. 

Our observations were not designed to target any particular type of star but only to give an unbiased survey of 24 \um excesses.  We are seeking primarily a statistical result that applies to a sample of field stars chosen without any prior knowledge of their stellar characteristics.  Explaining the individual origins of these infrared excesses requires more information for each star. It is especially crucial to know which stars are main sequence stars, because giants with excesses can be explained by dust having been produced and ejected in the extended stellar atmosphere.  

For determining the spectral classification of sources in Bo\"{o}tes, we initially utilized the DFBS, a low resolution objective prism survey of the sky originally used primarily to discover Markarian galaxies.   (The FLS field is not currently included within the DFBS.) These classifications provided the initial indication that the great majority of these stars are of F,G, and K spectral types.  

The DFBS cannot provide luminosity classifications, and the stars in Tables 1 and 2 are not sufficiently bright that other spectroscopic classifications have been made.  At present, our only way to estimate which stars are main sequence is to use available data on proper motions to estimate luminosity limits.  The proper motion data are from the Tycho-2 catalog \citep{hog00} (when available) or US Naval Observatory-B1.0 catalog \citep{mon03}.  Our estimate uses 50 km/s as an upper limit for $v_{t}$ to calculate the maximum distances and upper limits for luminosity.  For example, a star with proper motion  of 100 mas/yr should be at a distance less than 100 pc; for apparent $m_{v}$ = 15, this requires absolute magnitude $M_{v}$ $>$ 10.0.  Any star with $M_{v}$ $>$ 4 determined with this limit is assigned as main sequence (luminosity class V in Tables 1 and 2.)  This is a relaxed limit because $v_{t}$ may be smaller than the limit used, so many stars without an assignment to class V may also be main sequence stars. 

We also include in Tables 1 and 2 the optical $V$ magnitudes from the Tycho catalog or, as noted in the Tables, from the Hubble Space Telescope Guide Star Catalog, v. 2.3.2 \citep{mcL00} for the minority of stars not having Tycho measurements.  Using the $V$ magnitudes allows a consistent spectral classification for both Bo\"{o}tes and FLS fields, by assigning a spectral class using the $V$-$K$ colors of the stars compared to the relation between this color and spectral type.

We have determined an empirical transformation between ($V$-$K$) and spectral type using the stars in \citet{gor06} and \citet{car09} together with the M stars in \citet{gau07}.  This transformation is shown in Figure 2.  Spectral subclasses are not used after M0 because the relation is not well-calibrated; stars redder than ($V$-$K$) $>$ 4.5 are listed in the Tables only as M stars. 

As discussed in section 2.3, we use the criterion ($K$-[24]) $>$ 0.2 mag to select those stars with excess luminosity at 24 \um compared to the luminosity arising from the stellar photosphere.  \citet{gau07} show, however, that this criterion does not work for M stars because such stars are sufficiently cool that a color ($K$-[24]) $>$ 0.2 can arise just from the photosphere.  From Figure 1 in Gautier et al., a photospheric ($K$-[24]) $>$ 0.2 arises for stars with ($V$-$K$) $\ga$ 4.0; these are the M stars shown our Figure 2.  

\subsection{Fraction of FGK Main Sequence Stars with 24 \um Excesses}

In Figure 3 for Bo\"{o}tes stars and in Figure 4 for FLS stars, we show the distributions of ($V$-$K$) for the stars in Tables 1 and 2 and identify those 26 stars with excesses which are shown in Figure 1.  From the colors shown in Figures 3 and 4, we reject any M star with ($V$-$K$) $>$ 4.0 as having a real excess of luminosity above the photospheric luminosity.  This removes 4 stars from Bo\"{o}tes (stars 39, 76, 119, and 125 in Table 1) and 4 stars from the FLS (stars 55, 84, 89, and 136 in Table 2) from the list of stars whose ($K$-[24]) color can be taken as evidence of a real 24 \um excess.  The remaining 18 stars with excesses ($K$-[24]) $>$ 0.2 are all F, G, or K stars.  One of these 18 is a known variable star, source 107 in Table 1, so we do not include it as a candidate for a real excess.  We discuss only the remaining 17 stars and refer to these as the FGK stars with excesses. 

The upper limits on luminosities estimated from proper motions yield a lower limit for the number of FGK main sequence stars in Bo\"{o}tes and the FLS because other stars could also be main sequence stars without large proper motions.  This limit is 92 FGK main sequence stars in the combined samples in Tables 1 and 2.  Of the 17 FGK stars identified with 24 \um excesses, 8 of the 17 are main sequence based only on this limited criterion of proper motion, which gives a total fraction 8/92 (9\%) of main sequence FGK stars which have excesses.  

An independent estimate for which stars are main sequence stars derives by considering the magnitudes of the stars, as giants would be expected to appear systematically brighter.  This comparison is shown in Figures 3 and 4 for Bo\"{o}tes and the FLS.  Both diagrams show that most stars identified as having ($K$-[24]) $>$ 0.2 (open circles) are among the faintest stars of a given spectral type (i.e. of a given $V$-$K$ color).  In Figures 3 and 4 combined, only three stars with ($K$-[24]) $>$ 0.2  are above this trend, and these three are all M stars based on having ($V$-$K$) $>$ 4.0.  

All of the 17 FGK stars identified with 24 \um excesses in Bo\"{o}tes and the FLS fall on the lower envelope of $V$ mag. in Figures 3 and 4, implying that all 17 are main sequence stars; only 8 of these 17 are classified as main sequence stars in Tables 1 and 2 as determined from proper motions.  Adding these additional 9 FGK main sequence stars to the total of 92 FGK main sequence stars determined from proper motions yields a lower limit to the total number of FGK main sequence stars in the combined samples of Tables 1 and 2 of 101, of which 17 have 24 \um excesses.  The upper limit to the number of FGK main sequence stars is the entire sample of 141 FGK stars in Tables 1 and 2.  These results for estimating the main sequence stars based only on apparent magnitudes, therefore, yield limits to the fraction of main sequence stars with 24 \um excesses as somewhere between 17/141 (12\%) and 17/101 (17\%).  

Combining these limits derived from apparent magnitudes with the fraction of 9\% determined above using only proper motions, we can conclude that the fraction of main sequence FGK field stars with 24 \um excesses determined from comparison of 2MASS near-infrared and $Spitzer$ mid-infrared continuum flux densities is between 9\% and 17\%.  These results are similar to the 10\% fraction of main sequence stars identified with excesses from debris disks by \citet{mey08} and the 15\% fraction by \citet{car08}, so it is reasonable to conclude that most of the excesses we have identified arise from debris disks in main sequence stars.

\section{Summary and Conclusions}

Surveys at high galactic latitude at 24\,\um with the $Spitzer$ MIPS instrument are used in a serendipitous study of bright Galactic stars with detections at 24\,\ums.  268 stars with $V$ $\la$ 13 mag. and f$_{\nu}$(24 \ums) $>$ 1\,mJy are identified within 14 deg$^{2}$ of the combined NDWFS Bo\"{o}tes field and FLS field.  Using a comparison with $J$, $H$, and $K$ magnitudes from the 2MASS survey, it is found that 17 FGK stars  have excesses 0.2 $<$ ($K$-[24]) $<$ 0.7 compared to the expected ($K$-[24]) = 0.0 for stellar photospheres; i.e. the luminosity observed in the mid-infrared at 24 \um exceeds the photospheric luminosity by at least 20\%.

For main sequence stars, such excesses are attributed primarily to debris disks.  At least 8 of the 17 FGK stars with excesses are identified as main sequence stars based on proper motions; all 17 would be estimated as main sequence stars based only on distributions of apparent magnitude.  Depending on how many stars in the total sample really are main sequence stars, the fraction of main sequence field stars with excess luminosity at 24 \um in Bo\"{o}tes and the FLS is between 9\% and 17\%.  These results are comparable to the 7\% fraction for 69 bright FGK main sequence field stars in \citet{bry06} with 24 \um excess, to the 9\% fraction of 53 Pleiades FGK stars with excesses in \citet{gor06}, to the 10\% fraction of solar-like stars with excesses in \citet{mey08}, and the 15\% fraction in \citet{car08}. 

Within the uncertainties of the small number statistics, therefore, the Bo\"{o}tes and FLS surveys of Galactic stars find a similar fraction of main sequence stars with mid-infrared excesses as found among brighter, targeted field stars.  At present, however, a more quantitative result for Bo\"{o}tes and FLS stars requires additional observations. None of these stars have distance determinations through parallax measurements; the existing spectral classifications do not give luminosity classifications or determinations of variability; and we cannot rule out binary companions on a scale smaller than the resolution of the available images. 

Further observations of the individual stars are needed, therefore, to determine other possible sources of the identified excesses.  Our study primarily demonstrates the utility of combining $Spitzer$ surveys at 24\,\um with 2MASS data for bright stars to accumulate a meaningful sample of Galactic stars with apparent 24\,\um excesses.  Such unbiased, statistical samples of untargeted field stars should prove useful to eventual understanding of debris disks associated with solar-like stars and to discovering other explanations of mid-infrared excess luminosity.

\acknowledgments
We thank Lusine Sargsyan for help with data collection, and we thank M. Meyer for providing unpublished stellar coordinates.  We also thank the anonymous referee for detailed comments and helpful suggestions.  The $Spitzer$/MIPS Bo\"{o}tes survey was made possible with GTO time provided by the Infrared Spectrograph team and Marcia Rieke.  We thank NOAO for supporting the NOAO Deep Wide-Field Survey; AD and BJ acknowledge support from NOAO, which is operated by the Association of Universities for Research in Astronomy (AURA), Inc., under a cooperative agreement with the National Science  
Foundation.  This research has made use of the SIMBAD database,
operated at CDS, Strasbourg, France, and the NASA/IPAC Extragalactic Database (NED). This work is based on observations made with the
Spitzer Space Telescope, which is operated by the Jet Propulsion
Laboratory, California Institute of Technology under NASA contract
1407.  Support for this work by the IRS GTO team at Cornell University was provided by NASA through Contract Number 1257184 issued by JPL/Caltech. Support was also provided by the US Civilian Research and Development Foundation under grant ARP1-2849-YE-06.

\begin{deluxetable}{cccccccccc} 

\tablecolumns{10}
\tabletypesize{\footnotesize}


\tablewidth{0pc}
\tablecaption{Bright Stellar Infrared Sources in Bo\"{o}tes} 
\tablehead{

\colhead{\#} & \colhead{J2000 24\,\um coordinate} &\colhead{f$_{\nu}$(24 \ums)} & \colhead{2MASS $J$} & \colhead{2MASS $H$} & \colhead {2MASS $K$}& \colhead{Tycho $V$} &\colhead{($K$-[24])\tablenotemark{a}}& \colhead{Type\tablenotemark{b}}& \colhead{p.m.}\\ 
\colhead{}&  \colhead{} & \colhead{mJy} & \colhead{mag} & \colhead{mag} & \colhead{mag} & \colhead{mag} & \colhead{}& \colhead{} & \colhead{mas/yr}
}
\startdata

  1 & J142438.28+342521.2 &  10.43 &  7.71 &  7.25 &  7.12 &  9.63 &0.01 & K2V &   66   \\                          
  2 & J142438.60+342229.2 &   4.87 &  8.79 &  8.17 &  8.01 & 10.96 &0.07 & K5   &   10   \\                          
  3 & J142448.55+330528.1 &   3.84 &  8.51 &  8.24 &  8.20 &  9.74 &0.00 & G2  &9    \\                          
  4 & J142517.42+324727.4 &  17.47 &  6.98 &  6.62 &  6.52 &  8.47 &-0.03 & G6V  & 128  \\                          
  5 & J142524.04+332144.9 &   1.35 &  9.99 &  9.58 &  9.47 & 11.65 &0.14 & G9V &   95   \\                          
  6 & J142530.55+323740.5 &  24.91 &  6.98 &  6.35 &  6.20 &  9.17 &0.03 & K5    &   61   \\                          
  7 & J142537.43+325206.8 &   3.61 &  9.02 &  8.41 &  8.34 & 11.20 &0.08 & K4   &   2    \\                          
  8 & J142546.41+330430.4 &   1.30 &  9.56 &  9.36 &  9.38 & 10.47 &0.01 & F6  &   19   \\                          
  9 & J142548.78+324647.0 &   1.53 &  9.48 &  9.32 &  9.27 & 10.40 &0.07 & F7   &   30   \\                          
 10 & J142549.60+341533.5 &   1.21 & 10.04 &  9.56 &  9.49 & 11.83 &0.04 & K0  &   14   \\                          
 11 & J142550.61+340018.7 &   1.35 & 10.17 &  9.64 &  9.47 & 12.17\tablenotemark{f} &0.14 & K3   & \nodata       \\                          
 12\tablenotemark{c} & J142555.85+330426.4 &   1.75 &  9.98 &  9.39 &  9.28 & 11.91 &0.23 & K3  &   10   \\
 13\tablenotemark{c} & J142558.07+343133.5 &   1.58 & 10.38 &  9.86 &  9.65 & 13.36\tablenotemark{f} &0.49 & K9V  &   328  \\
 14 & J142559.84+342214.7 &   4.60 &  8.55 &  8.06 &  7.99 & 10.40 &-0.01 & K1V  &   329  \\                          
 15 & J142605.14+324951.8 &   1.37 &  9.84 &  9.50 &  9.43 & 11.38 &0.11 & G6   &   31   \\                          
 16 & J142630.15+344547.1 &   1.37 &  9.80 &  9.52 &  9.47 & 11.06 &0.15 & G2V  &   47   \\                          
 17 & J142631.81+321600.4 &   6.46 &  8.33 &  7.83 &  7.77 & 10.19 &0.14 & K1   &   8    \\                          
 18 & J142638.24+324740.9 &   4.77 &  8.13 &  7.92 &  7.92 &  9.13 &-0.04 & F8   &   49   \\                          
 19 & J142642.99+324335.7 &   5.17 &  8.98 &  8.22 &  8.02 & 11.85 &0.15 & K9   &   23   \\                          
 20 & J142651.59+340253.0 &   3.66 &  9.05 &  8.43 &  8.33 & 11.21 &0.08 & K4  &   6    \\                          
 21 & J142657.61+334323.3 &   1.31 &  9.99 &  9.52 &  9.44 & 11.91 &0.07 & K1   &   1    \\                          
 22 & J142659.23+332306.4 &   6.71 &  7.91 &  7.61 &  7.56 &  9.15 &-0.03 & G2   &   68   \\                          
 23 & J142703.45+330915.1 &   1.53 &  9.72 &  9.20 &  9.12 & 11.35 &0.08 & G9V  & 46   \\                          
 24 & J142709.48+331049.5 &  17.43 &  7.30 &  6.72 &  6.61 &  9.45 &0.05 & K4  & 19   \\                          
 25 & J142713.54+344319.9 &   6.86 &  7.78 &  7.64 &  7.59 &  8.65 &0.02 & F6   &   6    \\                          
 26 & J142719.49+341039.9 &   1.44 &  9.90 &  9.43 &  9.33 & 11.74 &0.07 & K1V  & 14   \\                          
 27 & J142743.08+330646.2 &   2.27 &  9.35 &  8.90 &  8.81 & 11.11 &0.04 & K0    &   44   \\                          
 28 & J142745.27+325703.0 &   4.55 &  8.10 &  8.01 &  7.99 &  8.67 &-0.02 & F0  &   24   \\                          
 29 & J142754.44+321611.4 &   1.99 &  9.53 &  9.07 &  9.01 & 11.27 &0.10 & G9  &   20   \\                          
 30 & J142758.17+353437.8 &  57.70 &  5.82 &  5.35 &  5.23 &  7.45 &-0.03 & G9V&   108  \\                          
 31 & J142801.62+321740.6 &   3.25 &  8.63 &  8.44 &  8.41 &  9.63 & 0.03& F8   &   11   \\                          
 32 & J142806.61+345307.5 &   1.16 & 10.06 &  9.63 &  9.54 & 11.84 & 0.04& K0   &   3    \\                          
 33 & J142807.08+345538.9 &   1.73 &  9.67 &  9.18 &  9.09 & 11.55 & 0.03& K1  &   34   \\                          
 34 & J142812.67+350259.8 &   1.22 &  9.90 &  9.51 &  9.44 & 10.92 &0.00 & G1V &   15   \\                          
 35 & J142838.88+322429.1 &   1.11 &  9.90 &  9.61 &  9.56 & 11.08 &0.02 & G2V  & 55   \\                          
 36 & J142840.78+330430.7 &   1.99 &  9.57 &  9.03 &  8.91 & 11.70 & 0.00& K4 &   17   \\                          
 37 & J142843.45+330354.1 &   8.46 &  8.25 &  7.55 &  7.40 & 10.53 &0.06 & K6   &   15   \\                          
 38 & J142845.57+343956.7 &   4.53 &  8.34 &  8.08 &  8.00 &  9.35 &-0.02 & F9  &15   \\                          
 39\tablenotemark{c} & J142850.46+350432.4 &  81.51 &  6.65 &  5.79 &  5.50 & 10.82 &0.62 &M   &   19   \\
 40\tablenotemark{d} & J142902.51+335038.6 &   9.26 &  7.85 &  7.47 &  7.36 &  9.31 & 0.12& G6V   &   75   \\
 41 & J142912.08+343054.4 &   1.39 & 10.35 &  9.63 &  9.43 & 12.66\tablenotemark{f} &0.13 & K7  & \nodata \\                 
 42 & J142916.52+321641.6 &   5.69 &  8.18 &  7.88 &  7.79 &  9.40 &0.02 & G3V &  30   \\                          
 43\tablenotemark{c} & J142918.28+333915.8 &   1.28 & 10.24 &  9.76 &  9.63 & 11.67 &0.24 & G7V  & 64   \\
 44 & J142922.59+334124.8 &   1.97 &  9.51 &  9.01 &  8.92 & 11.17\tablenotemark{f} &0.00 & G9 &  \nodata      \\                          
 45 & J142925.95+350319.1 &  10.85 &  7.75 &  7.24 &  7.10 &  9.67 &0.03 & K2V &  23   \\                          
 46 & J142929.72+331607.9 &   2.56 &  9.38 &  8.80 &  8.64 & 11.41 &0.00 & K4V&   60   \\                          
 47 & J142937.15+345856.0 &   1.01 &  9.98 &  9.76 &  9.74 & 10.89 &0.09 & F7  &   9    \\                          
 48 & J142939.11+331330.0 &   1.17 & 10.13 &  9.66 &  9.53 & 11.71 &0.04 & G9V & 36   \\                          
 49 & J142944.27+351031.4 &   1.35 &  9.97 &  9.47 &  9.36 & 11.89 &0.03 & K2V&   21   \\                          
 50 & J142959.60+353027.2 &   1.39 &  9.79 &  9.33 &  9.26 & 11.42 &-0.04 & G8  &   22   \\                          
 51 & J143001.31+354018.3 &   1.65 &  9.76 &  9.25 &  9.15 & 11.42 &0.04 & G9  &   36   \\                          
 52 & J143007.11+335605.6 &   3.48 &  8.61 &  8.34 &  8.31 &  9.59 &0.01 & F9V&   45   \\                          
 53 & J143007.34+334642.4 &  11.58 &  7.84 &  7.19 &  7.05 & 10.19 &0.05 & K6V&   28   \\                          
 54 & J143008.50+334803.7 &   1.04 &  9.87 &  9.67 &  9.65 & 10.72 &0.03 & F6V&   23   \\                          
 55 & J143010.72+322426.4 &   2.04 &  9.20 &  8.96 &  8.89 & 10.18 &0.01 & F9V&   48   \\                          
 56 & J143020.67+353121.8 &   4.23 &  8.36 &  8.06 &  8.06 &  9.42 &-0.03 & G0V&   43   \\                          
 57 & J143020.76+344359.2 &   2.49 &  9.07 &  8.74 &  8.68 & 10.39 &0.01 & G4V&   63   \\                          
 58 & J143032.35+345640.4 &  28.66 &  6.97 &  6.22 &  6.07 &  9.46 &0.05 & K7  &   15   \\                          
 59 & J143043.49+351806.5 &   3.54 &  8.58 &  8.31 &  8.24 &  9.84 &-0.05 & G2V&   130  \\                          
 60 & J143044.82+352714.3 &  25.05 &  6.76 &  6.31 &  6.14 &  8.82 &-0.02 & K3V&   521  \\                          
 61 & J143047.49+355541.5 &   2.04 &  9.31 &  8.98 &  8.93 & 10.44 &0.05 & G1  &   16   \\                          
 62 & J143053.52+345749.3 &   1.22 &  9.80 &  9.44 &  9.38 & 11.15 &-0.06 & G4  &   4    \\                          
 63 & J143055.56+351244.1 &   5.04 &  8.83 &  8.17 &  8.03 & 11.25 &0.13 & K7  &   9    \\                          
 64 & J143059.19+332213.9 &   1.33 &  9.95 &  9.48 &  9.38 & 11.46 &0.03 & G8V&   72   \\                          
 65 & J143101.97+325033.6 &   1.80 &  9.61 &  9.09 &  9.00 & 11.45 &-0.02 & K1  &   8    \\                          
 66 & J143102.12+350100.9 &   1.46 &  9.36 &  9.22 &  9.21 & 10.03 &-0.04 & F2  &   13   \\                          
 67 & J143104.48+322345.4 &   1.26 & 10.38 &  9.74 &  9.53 & 12.82\tablenotemark{f} &0.12 & K7V&   49   \\                          
 68 & J143104.54+342048.5 &   1.94 &  9.21 &  8.95 &  8.92 & 10.14 &-0.02 & F8  &   24   \\                          
 69 & J143105.00+355623.5 &   1.65 &  9.71 &  9.27 &  9.19 & 11.25 &0.08 & G7  &   30   \\                          
 70 & J143106.90+340935.2 &   1.21 &  9.66 &  9.40 &  9.36 & 10.75 &-0.09 & G0V&   59   \\                          
 71 & J143109.92+342532.3 &   3.09 &  8.76 &  8.46 &  8.47 &  9.98 &0.04 & G1  &   25   \\                          
 72 & J143116.97+331957.8 &   1.62 &  9.99 &  9.40 &  9.27 & 12.22 &0.14 & K5  &   16   \\                          
 73 & J143122.40+332034.7 &   1.20 &  9.79 &  9.54 &  9.53 & 10.95 &0.07 & G0V&   59   \\                          
 74 & J143123.54+330518.2 &  19.99 &  6.93 &  6.43 &  6.40 &  8.66 &-0.01 & G9V&   41   \\                          
 75 & J143126.60+341342.7 &   6.36 &  8.60 &  7.91 &  7.70 & 11.16 &0.05 & K8  &   2    \\                          
 76\tablenotemark{c} & J143143.11+321432.8 &   2.01 & 10.26 &  9.69 &  9.41 & 14.39\tablenotemark{f} &0.51 & MV&  197  \\
 77 & J143222.13+331137.9 &   2.47 &  8.60 &  8.64 &  8.59 &  8.69 &-0.09  &A1V&   9    \\                          
 78 & J143223.16+340938.4 &   4.40 &  8.88 &  8.27 &  8.11 & 11.21 &0.06  &K6  &   16   \\                          
 79 & J143230.69+344213.6 &   1.50 &  9.55 &  9.24 &  9.16 & 10.77 &-0.06  &G3V&   45   \\                          
 80 & J143231.54+344209.3 &   2.52 &  9.06 &  8.74 &  8.65 & 10.38 &0.00  &G4V&   46   \\                          
 81 & J143232.47+342328.1 &   1.22 &  9.86 &  9.58 &  9.52 & 10.91 & 0.08 &G0V &   25   \\                          
 82\tablenotemark{c} & J143249.14+334641.8 &   2.05 & 10.04 &  9.43 &  9.19 & 12.84\tablenotemark{f} &0.31  &K9V&   45   \\
 83 & J143253.74+332225.3 &   1.01 & 10.56 &  9.96 &  9.79 & 12.75\tablenotemark{f} &0.14  &K5  &  \nodata      \\                          
 84 & J143255.16+351540.6 &   1.41 &  9.90 &  9.41 &  9.31 & 11.69 & 0.02 &K0V&   73   \\                          
 85 & J143333.91+355153.4 &   2.44 &  9.29 &  8.80 &  8.67 & 11.11 &-0.02  &K1V&   22   \\                          
 86 & J143336.21+355802.6 &   2.60 &  \nodata  &  \nodata &  \nodata  &  \nodata &  \nodata & \nodata &\nodata        \\                          
 87\tablenotemark{e} & J143336.24+353509.0 &  19.79 &  6.31 &  5.96 &  5.89 &  7.63 &\nodata  &G4V&   201  \\
 88\tablenotemark{e} & J143336.39+353511.1 &  12.64 &  6.31 &  5.96 &  5.89 &  8.13 &\nodata  &G9V&   201  \\
 89 & J143337.80+332458.7 &   5.24 &  8.62 &  8.03 &  7.86 & 10.73 &0.00  &K4  &   19   \\                          
 90\tablenotemark{c} & J143346.08+333454.2 &   3.43 &  9.35 &  8.72 &  8.56 & 12.23 &0.24  &K9V&   40   \\
 91 & J143354.91+350735.8 &   1.06 & 10.06 &  9.75 &  9.66 & 11.41 &0.06  &G4V&   37   \\                          
 92 & J143357.18+334459.2 &   2.51 &  9.38 &  8.79 &  8.64 & 11.65 &-0.02  &K5V&   26   \\                          
 93 & J143401.78+341533.8 &   1.87 &  9.58 &  9.13 &  9.04 & 11.22 &0.06  &G9  &   13   \\                          
 94 & J143405.61+342543.2 &   1.22 & 10.09 &  9.60 &  9.47 & 12.38 &0.03  &K5  &   31   \\                          
 95 & J143407.17+331030.3 &   3.10 &  8.73 &  8.42 &  8.42 & 10.03 &-0.01  &G3V&   98   \\                          
 96 & J143418.44+342201.8 &   1.89 &  9.48 &  9.02 &  8.97 & 11.23 &0.00  &G9   &   32   \\                          
 97 & J143424.40+341818.7 &   2.92 &  8.82 &  8.60 &  8.56 &  9.85 &0.07  &F9   &   11   \\                          
 98 & J143428.44+332649.7 &   2.20 &  9.19 &  8.84 &  8.80 & 10.48 &0.00  &G3   &   55   \\                          
 99 & J143429.81+351730.9 &   1.28 & 10.17 &  9.59 &  9.42 & 12.25\tablenotemark{f} & 0.03 &K4  &   \nodata     \\                          
100 & J143447.78+353409.1 &   2.85 &  9.01 &  8.55 &  8.50 & 10.60 &-0.02  &G8V &   58   \\
101 & J143447.94+335010.1 &   1.08 & 10.28 &  9.69 &  9.55 & 11.99\tablenotemark{f} &-0.02  &K1   &  \nodata      \\                          
102 & J143457.58+334449.7 &  22.04 &  7.11 &  6.47 &  6.35 &  9.71 & 0.05 &K7V&   752  \\                          
103 & J143502.68+330437.4 &   1.23 & 10.27 &  9.70 &  9.55 & 12.38\tablenotemark{f} &0.12  &K4V&   738  \\                          
104 & J143508.74+331725.8 &   1.76 &  9.35 &  9.11 &  9.05 & 10.30 &0.01  &F8V&   41   \\                          
105 & J143516.38+354401.0 &   1.36 &  9.60 &  9.33 &  9.23 & 10.81 &-0.09  &G2V&   35   \\                          
106 & J143521.98+334855.3 &   3.75 &  8.62 &  8.30 &  8.17 &  9.91 & -0.05 &G4V&   148  \\                          
107\tablenotemark{c,d} & J143524.30+344646.7 &   7.76 &  9.57 &  8.86 &  8.62 & 12.25\tablenotemark{f} &1.19  &K9  &  \nodata      \\
108 & J143532.08+344112.2 &  35.35 &  6.53 &  5.94 &  5.80 &  8.67 &0.01  &K4  &   6    \\                          
109 & J143538.28+331141.0 &   2.07 &  9.55 &  9.07 &  8.91 & 11.23 &0.04  &K0  &   13   \\                          
110 & J143541.32+342744.2 &   1.29 &  9.82 &  9.46 &  9.37 & 11.18 &-0.01  &G5V&   78   \\                          
111 & J143554.98+350527.6 &  32.18 &  6.86 &  6.14 &  5.94 &  9.40 &0.05  &K8  &   17   \\                          
112 & J143618.58+341303.3 &   1.59 &  9.58 &  9.17 &  9.12 & 11.25 &-0.03  &G8V&   217  \\                          
113\tablenotemark{c} & J143621.58+344011.3 &   1.28 & 10.24 &  9.75 &  9.61 & 11.89\tablenotemark{f} &0.22  &K0  & \nodata       \\
114 & J143703.50+345209.2 &   2.39 &  8.99 &  8.80 &  8.70 & 10.05 &-0.01  &G0V&   37   \\                          
115 & J143705.75+334336.6 &  14.55 &  7.42 &  6.91 &  6.78 &  9.22 &0.03  &K1V&   22   \\                          
116 & J143707.34+334041.7 & 101.77 &  5.35 &  4.81 &  4.64 &  7.45 &0.00  &K4  &   35   \\                          
117 & J143725.45+343620.1 &   8.02 &  8.13 &  7.55 &  7.44 & 10.08 &0.04  & K3V&   31   \\                          
118\tablenotemark{c} & J143732.00+353950.6 &   1.84 & 10.29 &  9.78 &  9.66 & 12.24 &0.66  &K2V &   27   \\
119\tablenotemark{c} & J143732.02+352818.7 &  23.90 &  7.64 &  6.73 &  6.58 & 10.71 &0.37  &M0 &   13   \\
120\tablenotemark{c} & J143736.96+353940.6 &   1.14 & 10.20 & 10.00 &  9.94 & 11.29 &0.42  &F9  &   17   \\
121 & J143747.07+345204.1 &   3.15 &  8.68 &  8.48 &  8.42 &  9.61 &0.01  &F7V &   45   \\                          
122 & J143755.75+352035.9 &   1.07 & 10.18 &  9.67 &  9.54 & 11.86 &-0.04  &K0V &   26   \\                          
123 & J143808.21+334418.5 &   3.65 &  8.85 &  8.39 &  8.32 & 10.78 &0.07  &K1   &   15   \\                          
124 & J143821.43+345225.5 &   1.63 &  9.78 &  9.29 &  9.17 & 11.63 &0.04  &K1V &   11   \\                          
125\tablenotemark{c} & J143823.10+332929.8 &   2.62 &  9.81 &  9.19 &  8.98 & 12.94\tablenotemark{f} &0.37  &M0V  &   74   \\
126 & J143827.10+353906.3 &   2.25 &  9.69 &  9.01 &  8.94 & 11.76 &0.16  &K4   &   14   \\                          
127 & J143839.89+353113.0 &   5.05 &  8.61 &  8.10 &  7.97 & 10.47 &0.07  &K1  &   19   \\                          
128 & J143845.56+340208.2 &  19.93 &  7.11 &  6.59 &  6.42 &  9.18 & 0.01 &K3V &   31   \\                          

\enddata

\tablenotetext{a}{$K$-[24] color uses magnitude at 24\ums, [24], determined from f$_{\nu}$(24 \ums) with zero magnitude corresponding to f$_{\nu}$(24 \ums) = 7300 mJy.} 
\tablenotetext{b}{Spectral type is estimated from ($V$-$K$) color; luminosity class V (main sequence) is assigned for stars whose upper luminosity limit is $M_{v}$ $>$ 4, based on proper motions as discussed in text.} 
\tablenotetext{c}{Star identified in Fig. 1 as having 24 \um excess, with ($K$-[24]) $>$ 0.2 mag.}
\tablenotetext{d}{Variable star from \citet{ak00}.}
\tablenotetext{e}{Sources 87 and 88 are a close double star without individual $J$,$H$,$K$ magnitudes, spectral types, and proper motions.}
\tablenotetext{f}{$V$ mag. from Guide Star Catalog 2.3.2.}

\end{deluxetable}

\clearpage
\begin{deluxetable}{cccccccccc} 

\tablecolumns{10}
\tabletypesize{\footnotesize}

\tablewidth{0pc}
\tablecaption{Bright Stellar Infrared Sources in FLS} 
\tablehead{
\colhead{\#} & \colhead{J2000 24\,\um coordinate} &\colhead{f$_{\nu}$(24 \ums)} & \colhead{2MASS $J$} & \colhead{2MASS $H$} & \colhead {2MASS $K$}& \colhead{Tycho $V$} &\colhead{($K$-[24])\tablenotemark{a}}& \colhead{Type\tablenotemark{b}}& \colhead{p.m.}\\ 
\colhead{}&  \colhead{} & \colhead{mJy} & \colhead{mag} & \colhead{mag} & \colhead{mag} & \colhead{mag} &\colhead{}& \colhead{} & \colhead{mas/yr}
}

\startdata

  1 & J170756.92+594716.2 &   1.80 &  9.64 &  9.21 &  9.12 & 11.28 &0.10 & G8    &   26   \\                          
  2 & J170804.67+594359.4 &   3.58 &  8.98 &  8.44 &  8.34 & 10.86 &0.07 & K3   &   6    \\                          
  3 & J170804.74+593719.9 &   1.03 & 10.50 &  9.90 &  9.78 & 12.40\tablenotemark{d} &0.16 & K2    &  \nodata      \\                          
  4 & J170828.51+590529.7 &   1.08 & 10.35 &  9.82 &  9.71 & 12.08\tablenotemark{d} &0.14 & K0    &  \nodata      \\                          
  5 & J170904.61+585708.8 &   7.39 &  8.59 &  7.84 &  7.62 & 11.33 &0.13 & K9    &   9    \\                          
  6 & J170934.50+601428.5 &   4.83 &  8.67 &  8.10 &  8.00 & 10.60 &0.05 & K2    &   11   \\                          
  7 & J170934.58+591023.5 &   2.05 &  8.93 &  8.87 &  8.84 &  9.19 &-0.04 & A5    &   22   \\                          
  8 & J170938.29+583651.0 &   1.00 &  9.72 &  9.62 &  9.59 & 10.35 &-0.07 & F1    &   7    \\                          
  9 & J170944.84+600710.1 &   7.66 &  8.43 &  7.71 &  7.54 & 11.00 &0.09 & K8    &   7    \\                          
 10 & J170949.47+592011.5 &   2.03 &  9.64 &  9.12 &  8.99 & 11.66 &0.10 & K3&   10   \\                          
 11 & J170949.68+590334.2 &   2.85 &  8.81 &  8.64 &  8.56 &  9.65 & 0.04& F6   &   20   \\                          
 12 & J170954.31+602038.9 &   1.22 & 10.11 &  9.64 &  9.49 & 12.12 &0.05 & K2  &   9    \\                          
 13 & J171000.43+595221.0 &   1.60 &  9.76 &  9.31 &  9.22 & 11.51 &0.07 & K0    &   5    \\                          
 14 & J171028.15+594835.2 &   3.03 &  9.27 &  8.68 &  8.52 & 11.35 &0.06 & K4   &   4    \\                          
 15 & J171031.49+592603.7 &   1.26 &  9.83 &  9.53 &  9.46 & 11.05 &0.05 & G2   &   0    \\                          
 16 & J171043.26+595728.6 &   1.10 & 10.16 &  9.70 &  9.61 & 11.77\tablenotemark{d} &0.06 & G8V&   77   \\                          
 17 & J171046.34+584244.6 &   2.00 &  9.31 &  9.06 &  8.96 & 10.41 &0.04  &G1V  &   33   \\                          
 18 & J171049.11+592657.5 &   1.76 &  9.37 &  9.16 &  9.09 & 10.30 &0.05 & F8   &   12   \\                          
 19 & J171051.48+584608.0 &   6.40 &  8.38 &  7.82 &  7.68 & 10.35 &0.04  &K3   &   10   \\                          
 20 & J171054.36+600234.3 &   2.77 &  9.49 &  8.77 &  8.61 & 11.99 &0.06  &K7   &   \nodata     \\                          
 21\tablenotemark{c} & J171056.02+601821.2 &   1.05 & 10.87 & 10.04 &  9.85 & 13.61\tablenotemark{d} & 0.24 &K9   &  \nodata      \\
 22 & J171104.43+591343.9 &   1.69 &  9.34 &  9.14 &  9.10 & 10.35 &0.01  &F8   &   24   \\                          
 23 & J171117.62+590150.7 &   6.12 &  8.02 &  7.81 &  7.72 &  9.18 &0.03  &G1V&   19   \\                          
 24 & J171131.32+602556.3 &  15.78 &  7.77 &  7.01 &  6.78 & 10.77 & 0.12 &M0  &   15   \\                          
 25 & J171141.18+600502.2 &  14.15 &  7.83 &  7.06 &  6.92 & 10.54 &0.13  &K9  &   5    \\                          
 26 & J171142.54+590346.1 &   2.63 &  8.76 &  8.62 &  8.54 &  9.63 &-0.07  &F6   &   61   \\                          
 27 & J171146.27+584312.3 &   1.83 &  9.94 &  9.27 &  9.09 & 12.18\tablenotemark{d} &0.09  &K6   &  \nodata      \\                          
 28 & J171158.59+592633.9 &   1.44 &  9.66 &  9.38 &  9.29 & 11.02 &0.03  &G4   &   19   \\                          
 29 & J171201.59+602738.2 &   3.32 &  8.66 &  8.46 &  8.39 &  9.65 &0.03  &F8    &   16   \\
 30 & J171223.24+593259.5 &   2.19 &  9.48 &  8.93 &  8.88 & 11.59 &0.07  &K3  &   10   \\                          
 31 & J171228.09+583906.9 &   3.22 &  8.80 &  8.55 &  8.48 &  9.93 & 0.09 &G1V&   15   \\                          
 32 & J171244.56+590401.0 &  10.78 &  7.80 &  7.22 &  7.11 & 10.10 &0.03  &K5V&   481  \\                          
 33 & J171245.14+583631.7 &   1.16 &  9.94 &  9.61 &  9.57 & 11.14 & 0.07 &G2   &   12   \\                          
 34\tablenotemark{c} & J171256.67+583307.9 &   1.05 & 11.00 & 10.34 & 10.15 & 13.07\tablenotemark{d} &0.54  &K5   & \nodata       \\
 35 & J171303.68+594806.1 &   1.36 &  9.48 &  9.38 &  9.31 & 10.19 &-0.01  &F3   &   6    \\                          
 36 & J171308.55+591247.0 &   1.21 &  9.73 &  9.43 &  9.37 & 11.01 & -0.08 &G3   &   35   \\                          
 37 & J171315.84+585210.3 &   2.12 &  9.39 &  9.03 &  8.90 & 10.99 & 0.06 &G8V  &   40   \\                          
 38 & J171316.32+584005.9 &   4.23 &  8.85 &  8.37 &  8.21 & 10.91 &0.12  &K3V &   31   \\                          
 39 & J171319.03+602604.7 &   1.88 &  9.27 &  9.09 &  9.06 &  9.99 & 0.09 &F4   &   16   \\                          
 40 & J171325.70+602347.1 &   1.05 &  9.95 &  9.62 &  9.58 & 11.40 &-0.03  &G5V &   31   \\                          
 41 & J171330.81+593842.7 &   1.13 &  9.88 &  9.59 &  9.50 & 11.25 & -0.03 &G4V &   24   \\                          
 42 & J171340.08+584628.6 &   1.47 &  9.47 &  8.97 &  8.87 & 11.32 & -0.37 &K1   &   34   \\                          
 43 & J171403.61+590843.9 &   5.12 &  8.56 &  8.07 &  7.96 & 10.48 &0.07  &K2   &   14   \\                          
 44 & J171428.56+602524.8 &   1.50 &  9.91 &  9.37 &  9.28 & 11.45 &0.06  &G9V&   82   \\                          
 45\tablenotemark{c} & J171510.58+602003.5 &   1.03 & 10.14 &  9.94 &  9.87 & 11.26 &0.24  &G0V   &   26   \\
 46 & J171512.67+591203.3 &  43.73 &  6.49 &  5.80 &  5.62 &  9.06 &0.06  &K8   &   20   \\                          
 47 & J171514.82+593331.8 &   2.47 &  9.61 &  8.98 &  8.80 & 11.80\tablenotemark{d} &0.12  &K5   &  \nodata      \\                          
 48 & J171555.80+593024.4 &  15.31 &  7.75 &  7.02 &  6.81 & 10.75 &0.11  &M0V   &   18   \\                          
 49 & J171556.73+594014.8 &   1.32 & 10.10 &  9.53 &  9.39 & 12.03 & 0.03 &K3   &  \nodata      \\                          
 50 & J171617.80+600926.9 &   3.48 &  8.97 &  8.51 &  8.37 & 10.70 & 0.07 &K0    &   10   \\                          
 51 & J171649.32+594803.3 &   1.52 & 10.16 &  9.49 &  9.31 & 12.64\tablenotemark{d} &0.11  &K7V &   146  \\                          
 52 & J171650.79+594914.5 &   1.21 & 10.52 &  9.85 &  9.65 & 13.15\tablenotemark{d} &0.20 &K8V &   150  \\                          
 53 & J171659.33+593156.5 &   4.33 &  8.76 &  8.25 &  8.14 & 10.71 & 0.07 &K2V &   12   \\                          
 54 & J171707.56+600528.2 &   1.10 & 10.18 &  9.73 &  9.64 & 11.92 &0.09  &K0V&   70   \\                          
 55\tablenotemark{c} & J171709.8+592409.3 &   1.06 & 10.85 & 10.31 & 10.06 & 14.43\tablenotemark{c} &0.46  &M0V&   301  \\
 56 & J171731.38+593525.1 &   3.54 &  9.37 &  8.60 &  8.41 & 11.81\tablenotemark{d} &0.12  &K7   &  \nodata      \\                          
 57 & J171732.62+592523.4 &   2.32 &  9.16 &  8.83 &  8.73 & 10.60 &-0.01  &G6V &   167  \\                          
 58 & J171742.76+584458.2 &   2.16 &  9.18 &  8.95 &  8.91 & 10.18 &0.09  &F8   &   12   \\                          
 59 & J171756.24+601026.6 &   7.97 &  8.27 &  7.64 &  7.50 & 10.66 &0.10  &K6   &   9    \\                          
 60 & J171831.13+590316.7 &   1.26 &  9.85 &  9.60 &  9.55 & 10.99 &0.14  &G0V &   59   \\                          
 61 & J171835.04+593055.4 &   5.97 &  8.44 &  7.92 &  7.83 & 10.53 &0.11  &K3   &   5    \\                          
 62 & J171840.32+585221.9 &   1.54 &  9.55 &  9.36 &  9.28 & 10.48 &0.09  &F8V &   31   \\                          
 63 & J171841.73+600247.3 &   1.05 & 10.31 &  9.85 &  9.76 & 12.26 & 0.15 &K1   &   4    \\                          
 64 & J171848.39+585535.8 &   1.01 & 10.53 &  9.94 &  9.79 & 12.56\tablenotemark{d} &0.14  &K4   &  \nodata      \\                          
 65 & J171909.21+593234.0 &   1.69 &  9.44 &  9.16 &  9.07 & 10.61 & -0.02 &G2V &   35   \\                          
 66 & J171914.85+583610.1 &   1.18 & 10.05 &  9.41 &  9.22 & 12.36\tablenotemark{d} &-0.25  &K6  &   \nodata     \\                          
 67 & J171930.72+592802.3 &   1.68 &  9.85 &  9.34 &  9.16 & 11.85 &0.06  &K3   &   25   \\                          
 68 & J171931.85+583145.1 &   6.48 &  7.86 &  7.69 &  7.65 &  9.34 &0.02  &G3  &   14   \\                          
 69 & J171942.92+592311.2 &   1.11 &  9.84 &  9.58 &  9.53 & 11.03 &-0.01  &G1  &   20   \\                          
 70 & J171959.97+583612.3 &   3.54 &  9.18 &  8.54 &  8.40 & 11.29 &0.11  &K4V&   16   \\                          
 71 & J172001.70+593938.1 &   1.60 &  9.46 &  9.21 &  9.14 & 10.59 & -0.01 &G1  &   16   \\                          
 72 & J172002.12+590623.3 &   1.43 &  9.60 &  9.46 &  9.39 & 10.28 & 0.12 &F3   &   21   \\                          
 73 & J172015.42+583838.6 &   1.69 &  9.97 &  9.34 &  9.20 & 12.03\tablenotemark{d} & 0.11 &K4   &  \nodata      \\                          
 74 & J172027.26+591156.3 &   4.95 &  7.98 &  7.94 &  7.92 &  8.35 & 0.00 &A6   &   14   \\                          
 75 & J172029.13+584812.5 &   9.94 &  8.36 &  7.57 &  7.32 & 11.76 &0.16  &M1   &   10   \\                          
 76\tablenotemark{c} & J172120.81+600329.3 &   1.96 &  9.81 &  9.30 &  9.17 & 11.76 &0.24  &K2  &   13   \\
 77 & J172127.65+595120.7 &   4.00 &  8.84 &  8.30 &  8.17 & 10.78 &0.02  &K2   &   10   \\                          
 78 & J172129.38+585428.0 &   1.22 &  9.98 &  9.59 &  9.54 & 11.43 &0.10  &G6   &   9    \\                          
 79 & J172133.87+601359.3 &   1.09 & 10.26 &  9.84 &  9.72 & 11.89\tablenotemark{d} &0.16  &G9V&   93   \\                          
 80\tablenotemark{c} & J172149.98+602606.2 &   2.19 &  9.70 &  9.16 &  9.02 & 11.52 &0.21  &K1   &   8    \\
 81 & J172159.80+593114.2 &   9.43 &  8.00 &  7.35 &  7.29 & 10.04 & 0.07 &K3   &   21   \\                          
 82 & J172205.65+582723.6 &   1.00 & 10.21 &  9.89 &  9.84 & 11.48 &0.18  &G3V  &   39   \\                          
 83 & J172212.47+590054.3 &   9.89 &  7.94 &  7.39 &  7.26 & 10.21 &0.09  &K5    &   7    \\                          
 84\tablenotemark{c} & J172218.32+595003.8 &   1.51 & 10.58 &  9.99 &  9.70 & 14.55\tablenotemark{d} &0.49  &MV   &   109  \\
 85 & J172222.48+583320.5 &   1.15 & 10.31 &  9.71 &  9.62 & 12.26\tablenotemark{d} &0.11  &K3V &   106  \\                          
 86 & J172227.48+595432.6 &   1.69 &  9.85 &  9.23 &  9.10 & 12.13 &0.01  &K5V   &   14   \\                          
 87 & J172237.69+602046.4 &   1.52 &  9.65 &  9.28 &  9.19 & 11.18 &-0.01  &G7V &   87   \\                          
 88 & J172237.85+591109.6 &   2.42 &  9.22 &  8.81 &  8.75 & 10.95 &0.05  &G9   &   10   \\                          
 89\tablenotemark{c} & J172242.86+600822.5 &   2.32 & 10.05 &  9.42 &  9.16 & 14.29\tablenotemark{d} &0.42  &MV   &   334  \\
 90 & J172244.74+582612.9 &  10.20 &  7.84 &  7.33 &  7.21 &  9.76 &0.07  &K2   &   14   \\                          
 91 & J172247.80+600800.0 &   2.22 &  9.19 &  8.91 &  8.83 & 10.55 &0.04  &G4   &   49   \\                          
 92 & J172248.83+591015.4 &   2.58 &  9.60 &  8.91 &  8.77 & 11.89 &0.14  &K6  &   4    \\                          
 93 & J172253.01+601903.6 &   4.06 &  8.75 &  8.27 &  8.18 & 10.50 &0.04  &K0V   &   51   \\                          
 94 & J172305.30+602555.0 &   6.94 &  8.62 &  7.85 &  7.68 & 11.34 &0.13  &K9  &   10   \\                          
 95 & J172309.22+591905.2 &   1.96 &  9.17 &  9.07 &  9.01 &  9.77 &0.08  &F1  &   10   \\                          
 96 & J172316.20+582824.1 &   1.19 & 10.35 &  9.79 &  9.67 & 12.20\tablenotemark{d} & 0.20 &K2   &  \nodata      \\                          
 97 & J172338.57+583418.4 &   3.29 &  9.09 &  8.58 &  8.48 & 11.36 &0.11  &K4V &   17   \\                          
 98 & J172346.48+593432.2 &   9.25 &  8.08 &  7.48 &  7.34 & 10.43 &0.10  &K6   &   6    \\                          
 99 & J172349.35+585637.9 &   1.37 &  9.65 &  9.38 &  9.35 & 10.51 &0.03  &F7  &   15   \\                          
100 & J172353.29+594657.9 &   2.90 &  8.75 &  8.56 &  8.51 &  9.73 &0.01  &F8   &   33   \\                          
101 & J172354.00+592047.7 &   7.67 &  8.04 &  7.63 &  7.51 &  9.64 &0.06  &G8   &   21   \\                          
102 & J172356.40+601334.3 &   1.37 &  9.64 &  9.39 &  9.34 & 10.81 &0.02  &G4   &   13   \\                          
103 & J172404.49+602404.4 &  21.23 &  7.26 &  6.60 &  6.45 &  9.59 &0.11  &K6   &   6    \\                          
104 & J172412.06+592110.0 &  16.47 &  7.58 &  6.87 &  6.71 &  9.98 &0.09  &K7    &   17   \\                          
105 & J172413.29+584155.2 &   4.66 &  8.19 &  7.98 &  7.97 &  8.98 &-0.02  &F5   &   9    \\                          
106 & J172419.58+601335.3 &   1.24 &  9.79 &  9.54 &  9.47 & 10.96 & 0.05 &G1   &   12   \\                          
107 & J172428.78+590913.4 &   7.16 &  8.20 &  7.68 &  7.58 & 10.11 &0.06  &K2   &   2    \\                          
108 & J172448.22+600454.4 &   3.57 &  9.51 &  8.53 &  8.36 & 11.58\tablenotemark{d} & 0.08 &K7   &  \nodata      \\                          
109 & J172452.46+595915.7 &   1.23 & 10.24 &  9.72 &  9.61 & 12.70 &0.18  &K6  &   17   \\                          
110 & J172458.95+584728.6 &   3.03 &  8.81 &  8.57 &  8.51 &  9.86 & 0.06 &F9  &   13   \\                          
111 & J172514.49+595525.3 &   8.77 &  8.16 &  7.56 &  7.38 & 10.43 &0.08  &K0    &   8    \\                          
112\tablenotemark{c} & J172515.33+583508.2 &   1.22 & 10.02 &  9.69 &  9.67 & 11.21\tablenotemark{d} &0.23  &G2  & \nodata       \\
113\tablenotemark{c} & J172540.84+591531.2 &   1.31 & 10.01 &  9.77 &  9.71 & 10.51 & 0.34 &F2   &   13   \\
114 & J172541.34+600254.2 &  54.75 &  5.49 &  5.53 &  5.50 &  5.65 &0.19  &A2V&   26   \\                          
115\tablenotemark{c} & J172549.45+583347.6 &   1.10 & 10.43 &  9.92 &  9.81 & 12.13\tablenotemark{d} &0.26  &K0   &   \nodata     \\
116 & J172554.16+594319.8 &   1.29 &  9.74 &  9.53 &  9.47 & 10.77 & 0.09 &F9  &   11   \\                          
117 & J172604.83+583907.0 &  24.94 &  6.17 &  6.16 &  6.15 &  6.51 &-0.02  &A5V &   17   \\                          
118 & J172608.49+585404.7 &   1.17 & 10.08 &  9.59 &  9.50 & 11.69\tablenotemark{d} &0.01  &G9  &  \nodata      \\                          
119 & J172611.34+591503.3 &   1.61 &  9.68 &  9.27 &  9.19 & 11.19 &0.05  &G7V&   90   \\                          
120 & J172616.32+593204.5 &   1.26 &  9.97 &  9.49 &  9.39 & 12.15 &-0.02  &K3  &   9    \\                          
121 & J172616.49+590105.2 &   1.51 &  9.94 &  9.39 &  9.25 & 11.32 &0.04  &G7  &   7    \\                          
122 & J172619.25+601749.1 &  25.75 &  7.16 &  6.45 &  \nodata     &  9.93 &\nodata  &K2    &   13   \\                          
123 & J172624.94+601735.3 &   5.15 &  9.02 &  8.21 &  8.02 & 12.13 &0.14  &M0  &  \nodata      \\                          
124 & J172627.66+600212.7 &   2.16 &  9.48 &  8.94 &  8.83 & 11.42 &0.01  &K2 &   8    \\                          
125 & J172628.19+590217.9 &  50.00 &  6.14 &  5.63 &  5.47 &  8.10 &0.06  &K3   &   8    \\                          
126 & J172636.15+595844.2 &   1.08 & 10.09 &  9.60 &  9.51 & 11.95 &-0.06  &K1V &   19   \\                          
127 & J172642.09+593809.2 &   1.75 &  9.83 &  9.28 &  9.14 & 11.96 &0.09  &K4  &   14   \\                          
128 & J172655.20+601838.0 &   1.00 &  9.92 &  9.72 &  9.68 & 10.92 &0.02  &F8  &   8    \\                          
129 & J172655.93+590000.2 &   1.31 &  9.70 &  9.53 &  9.44 & 10.72 &0.07  &F9   &   4    \\                          
130 & J172656.45+583358.2 &   3.43 &  9.23 &  8.53 &  8.37 & 11.52 &0.05  &K6  &   9    \\                          
131 & J172659.82+582754.8 &   8.60 &  8.50 &  7.68 &  7.49 & 11.27 &0.17  &K9  &   8    \\                          
132 & J172701.01+595415.8 &   2.62 &  8.99 &  8.67 &  8.62 & 10.35 & 0.01 &G4   &   43   \\                          
133 & J172710.30+601050.2 &   3.45 &  9.01 &  8.48 &  8.37 & 10.63 &0.06  &G9  &   5    \\                          
134 & J172714.55+593338.4 &   2.64 &  8.97 &  8.61 &  8.50 & 10.32 &-0.10  &G5V   &   26   \\                          
135 & J172722.36+594712.9 &   2.19 &  9.20 &  8.94 &  8.86 & 10.40 &0.05  &G2V&   46   \\                          
136\tablenotemark{c} & J172730.49+594449.6 &  18.91 &  7.72 &  6.88 &  6.68 & 10.98 &0.21  &M0  &   10   \\
137 & J172804.14+593648.1 &   1.38 &  9.72 &  9.44 &  9.37 & 11.10 &0.06  &G4V &   15   \\                          
138\tablenotemark{c} & J172823.53+601924.8 &   1.11 & 10.72 & 10.08 &  9.86 & 13.74\tablenotemark{d} &0.31  &K9V &   89   \\
139 & J172826.02+591257.7 &   7.77 &  8.26 &  7.65 &  7.51 & 10.61 & 0.08 &K6   &   5    \\                          
140 & J172903.46+592657.9 &   5.36 &  8.80 &  8.16 &  7.98 & 11.12 &0.14  &K6V &   13   \\                          
                                                                 
\enddata

\tablenotetext{a}{$K$-[24] color uses magnitude at 24 \ums, [24], determined from f$_{\nu}$(24 \ums) with zero magnitude corresponding to f$_{\nu}$(24 \ums) = 7300 mJy.}
\tablenotetext{b}{Spectral type is estimated from ($V$-$K$) color; luminosity class V (main sequence) is assigned for stars whose upper luminosity limit is $M_{v}$ $>$ 4, based on proper motions as discussed in text.} 
\tablenotetext{c}{Star identified in Fig. 1 as having 24 \um excess, with ($K$-[24]) $>$ 0.2 mag.}
\tablenotetext{d}{$V$ mag. from Guide Star Catalog 2.3.2.}

\end{deluxetable}                                                             
\clearpage

\begin{figure}
\figurenum{1}
\includegraphics[scale=1.5]{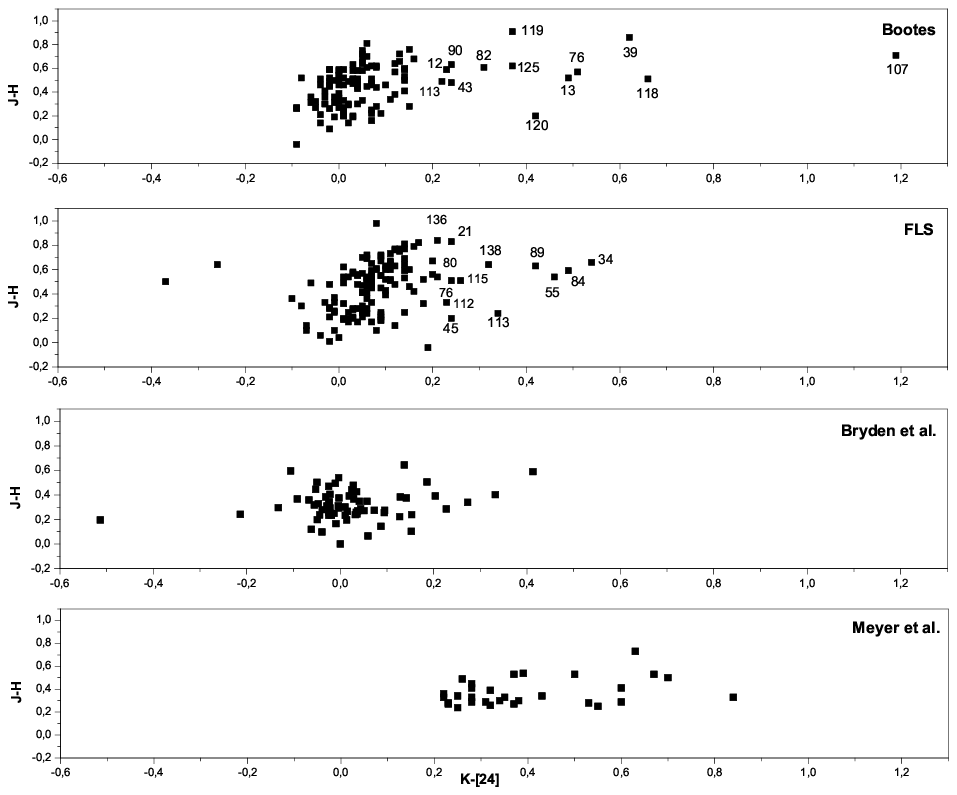}
\caption{Upper panel: color-color diagram for all Bo\"{o}tes sources in Table 1 identifying sources with ($K$-[24]) $>$ 0.2 mag.; 
second panel: same diagram for all FLS sources in Table 2; third panel: same diagram for all field stars in \citet{bry06}; bottom panel: same diagram for all stars with 24 \um excesses in \citet{mey08}. Magnitude at 24\,\um is defined as [24], with zero magnitude of the MIPS 24\,\um fluxes corresponding to 7300 mJy. Stars with spectral types earlier than M0 with 24 \um luminosity arising only from the stellar photosphere should have ($K$-[24]) = 0. }
\end{figure}

\begin{figure}
\figurenum{2}
\includegraphics[scale=1.5]{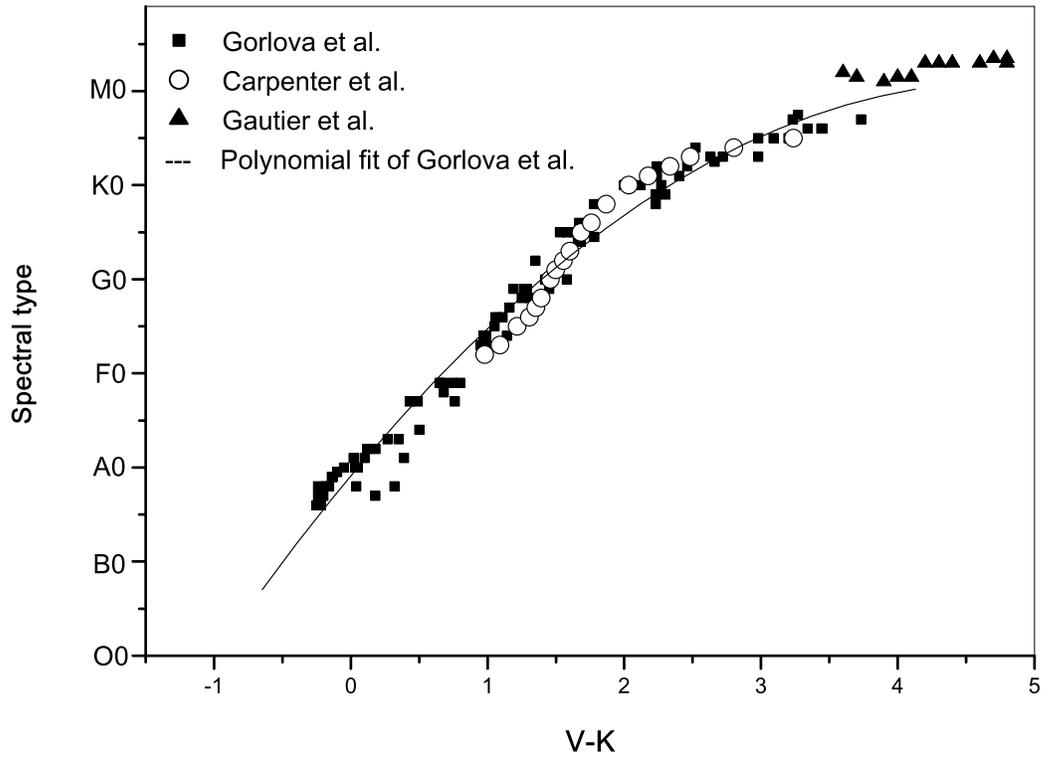}
\caption{Solid curve shows transformation used to convert ($V$-$K$) color to spectral classes for Tables 1 and 2; data points shown derive from references given.}
\end{figure}

\begin{figure}
\figurenum{3}
\includegraphics[scale=1.5]{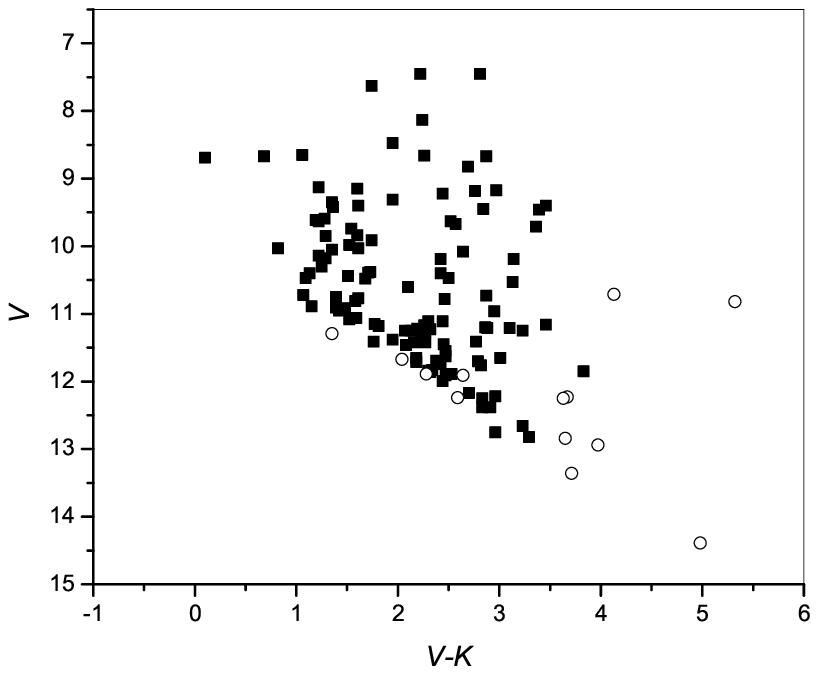}
\caption{Distribution of $V$ magnitude compared to $V$-$K$ color for Bo\"{o}tes stars in Table 1. Open circles identify objects with infrared color ($K$-[24]) $>$ 0.2 in Figure 1. Range of color arises from range of spectral types included as shown in Figure 2;  A0 has $V$-$K$ = 0, and M0 has $V$-$K$ $\sim$ 4.0. Stars with $V$-$K$ $>$ 4.0 can have ($K$-[24]) $>$ 0.2 arising only from their photospheres, so these stars are not considered to have real 24 \um excesses. } 
\end{figure}

\begin{figure}
\figurenum{4}
\includegraphics[scale=1.5]{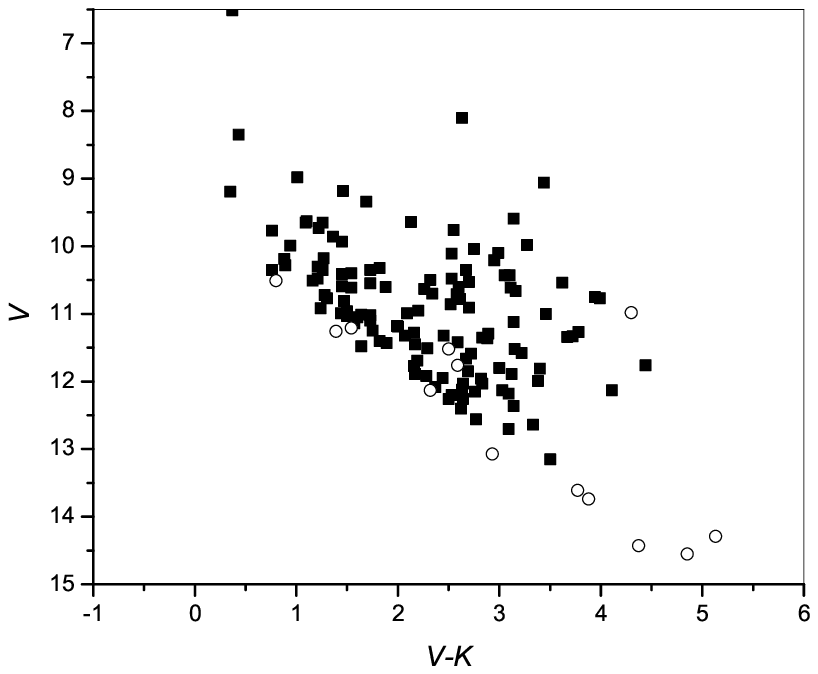}
\caption{Distribution of $V$ magnitude compared to $V$-$K$ color for FLS stars in Table 2. Open circles identify objects with infrared color ($K$-[24]) $>$ 0.2 in Figure 1. Range of color arises from range of spectral types included as shown in Figure 2;  A0 has $V$-$K$ = 0, and M0 has $V$-$K$ $\sim$ 4.0. Stars with $V$-$K$ $>$ 4.0 can have ($K$-[24]) $>$ 0.2 arising only from their photospheres, so these stars are not considered to have real 24 \um excesses.} 
\end{figure}

\end{document}